\documentclass[epsf,twocolumn,showpacs,preprintnumbers]{revtex4}
\usepackage{graphics}
\usepackage{graphicx}
\usepackage{dcolumn} 
\usepackage{bm}
\usepackage{epsfig}
\begin{document}
\title{Superconductivity and Lattice Instability in Compressed Lithium
from Fermi Surface Hot Spots} 
\author{Deepa Kasinathan,$^1$ J. Kune\v{s},$^{1,2}$ A. Lazicki,$^{1,3}$ 
  H. Rosner,$^4$
  C. S. Yoo,$^3$ R. T. Scalettar,$^1$ and W. E. Pickett$^1$} 
\affiliation{$^1$Department of Physics, University of California Davis, 
  Davis, CA 95616}
\affiliation{$^2$Institute of Physics, ASCR, Cukrovarnick\'a 10,
  162 53 Praha 6, Czech Republic}
\affiliation{$^3$Lawrence Livermore National Laboratory, Livermore
  CA 94551} 
\affiliation{$^4$Max-Planck-Instit\"ut f\"ur Chemische Physik 
  fester Stoffe Dresden, Germany}
\date{\today}
\begin{abstract}
The highest superconducting temperature T$_c$ observed in any elemental
metal (Li with T$_c \sim$ 20 K at pressure P$\sim$ 40 GPa) 
is shown to arise from critical (formally divergent)
electron-phonon coupling to the transverse 
${\cal T}_1$ phonon branch along
intersections of {\it Kohn anomaly surfaces} with the Fermi surface.
First principles linear response
calculations of the phonon spectrum and spectral function
$\alpha^2 F(\omega)$ reveal (harmonic) instability 
already at 25 GPa. 
Our results imply that the fcc phase is anharmonically stabilized
in the 25-38 GPa range.
\end{abstract}

\maketitle

The recent observations of superconductivity in fcc Li up to T$_c$ = 14 K
in near-hydrostatic fcc-phase samples,\cite{schilling} 
and as high as 20 K in
non-hydrostatic pressure cells,\cite{shimizu,struzhkin} in the pressure 
range 20 GPa $\leq$P$\leq$ 40 GPa 
provides almost as startling a development as the 
discovery\cite{akimitsu} in 2001
of T$_c$ = 40 K in MgB$_2$.  Lithium at ambient conditions, 
after all, is a simple $s$-electron metal
showing no superconductivity above 100 $\mu$K.\cite{finns} 
What can possibly transform it into the best elemental 
superconductor known, still in a simple, monatomic, cubic phase?
There is no reason to suspect a magnetic 
(or other unconventional) pairing 
mechanism, but it seems equally unlikely that it transforms into
a very strongly coupled electron-phonon (EP) superconductor 
at readily accessible pressures.

The strength of EP coupling in Li 
has attracted attention for some
time.  Evaluations based on empirical pseudopotentials\cite{allenLi}
early on suggested substantial coupling strength $\lambda$=0.56
and hence readily observable superconductivity (T$_c >$ 1 K); more recent
calculations relying on the rigid muffin-tin approximation (RMTA) 
reached a similar conclusion\cite{jarlborg,novikov} 
and led to prediction of remarkably high
T$_c \sim 70$ K under pressure.\cite{novikov}   
None of these studies actually 
calculated phonon frequencies, relying instead on estimates of a
representative phonon frequency ${\bar \omega}$
based on the Debye temperature,
which is only an extrapolation from the $q\rightarrow 0$ phonons.
Linear response calculations of the phonons and 
EP coupling\cite{liu} in bcc
Li confirmed that superconductivity would occur in 
bcc Li ($\lambda$ = 0.45), but superconductivity is not observed due to
the transformation into the 9R phase with 25\% weaker
coupling.
Experimentally, superconductivity only appears above 20 GPa in
the fcc phase.

In this paper we focus on the monatomic fcc phase that is stable in the
20-38 GPa range.  After providing additional characterization of the
previously discussed\cite{neaton,iyakutti,hanfland,rodriguez}
evolution of the electronic structure under pressure, 
we analyze the implications of the Fermi surface (FS) topology for
properties of Li.
To study $\lambda$ 
microscopically we focus on
the decomposition\cite{allen} into mode 
coupling strengths $\lambda_{Q\nu}$, where
$\lambda = (1/3N)\sum_{Q\nu} \lambda_{Q\nu} = <\lambda_{Q\nu}>$ is the
Brillouin zone (BZ) and
phonon branch ($\nu$) average.  We find that increase of pressure leads to
{\it very strong} EP coupling to a {\it specific branch} in 
{\it very restricted regions} of momentum space determined by the
FS topology;
these features are directly analogous to 
the focusing of coupling strength\cite{ucd,kortus,kong} in MgB$_2$.  
Unlike in MgB$_2$,
tuning with pressure leads to a vanishing
harmonic frequency at $\sim$25 GPa, beyond which the fcc phase 
is stabilized by anharmonic interactions.

\begin{figure}
\rotatebox{-00}{\resizebox{5.5cm}
{5.5cm}
    {\includegraphics{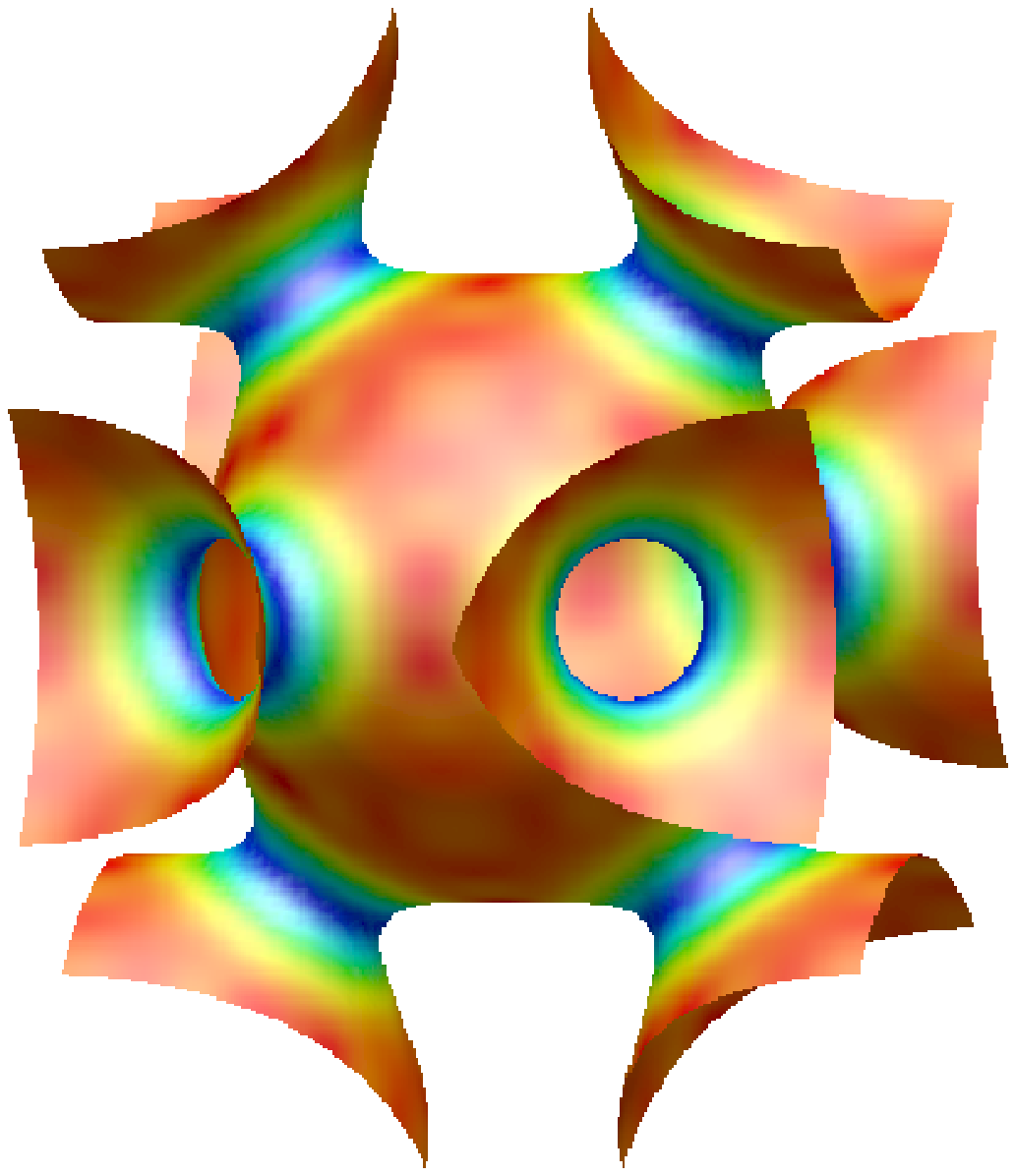}}}
\rotatebox{-00}{\resizebox{5.5cm}{4.8cm}
   {\includegraphics{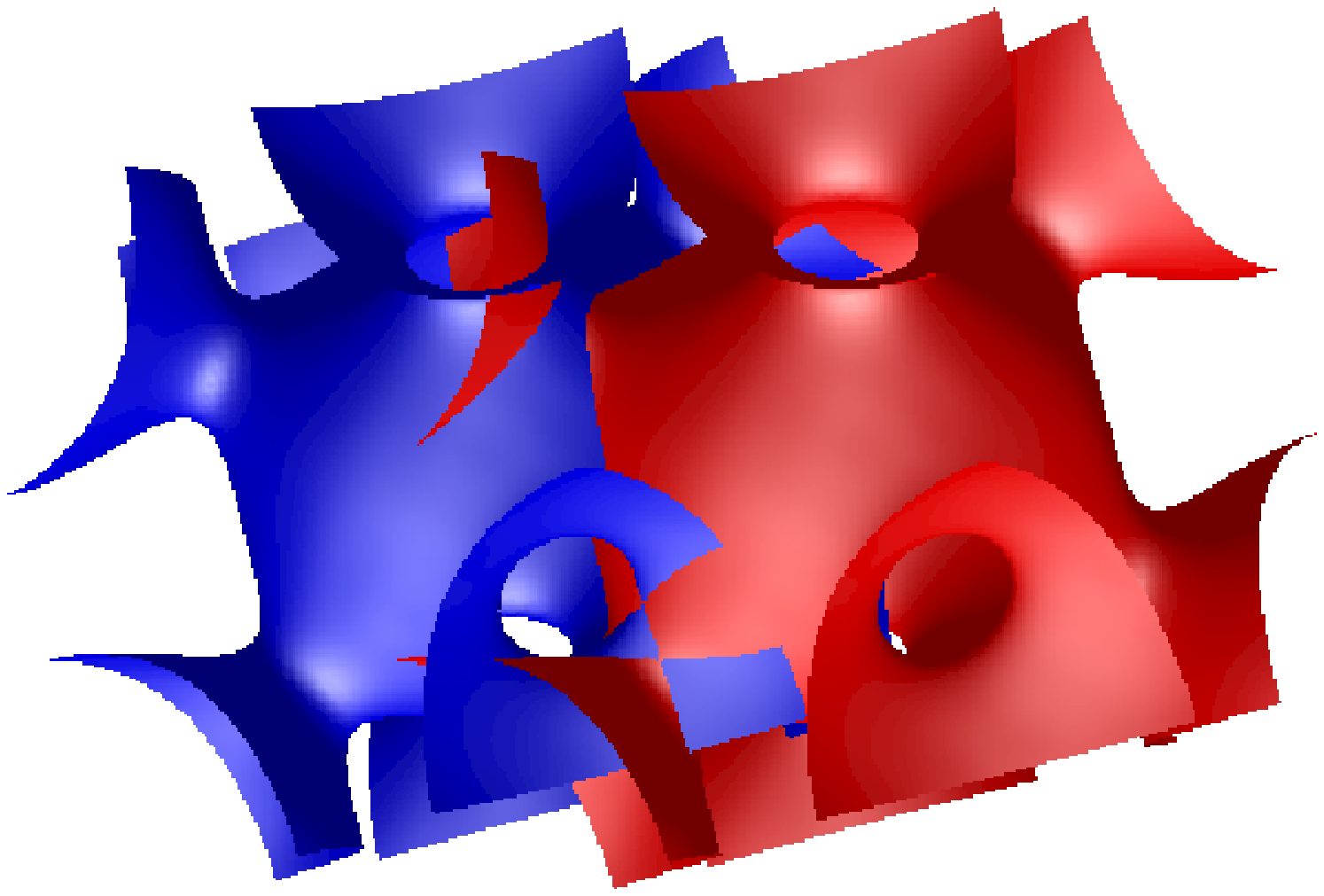}}}
\vskip -2mm
\rotatebox{-90}{\resizebox{5.5cm}{5.5cm}
   {\includegraphics{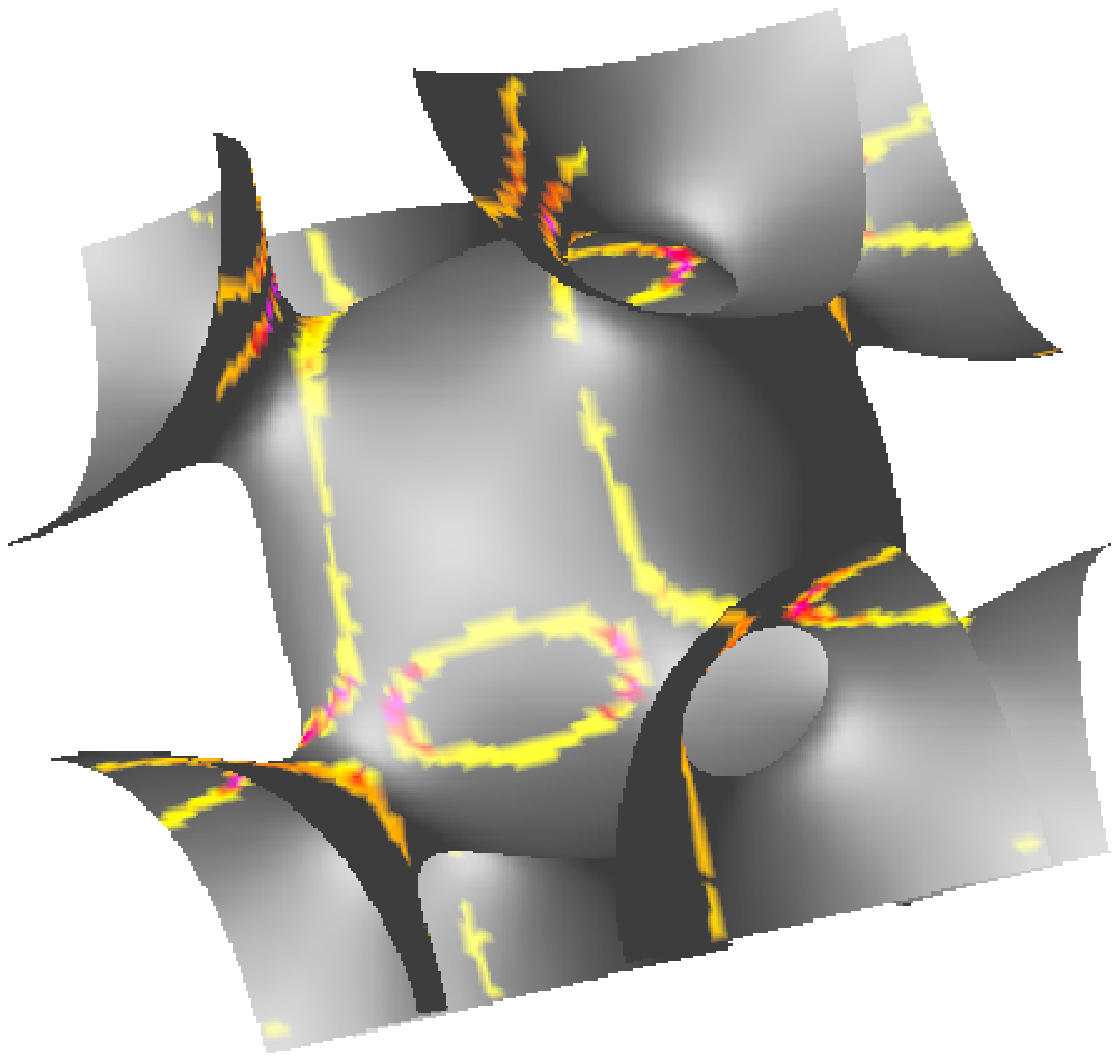}}}
\caption{(color online) {\it Top figure}: Fermi surface of Li at 35 GPa plotted in a
cube region around k=0 and colored by
the value of the Fermi velocity.  Red (belly areas) denotes fast electrons
($v_F^{max}$ = 9$\times 10^7$ cm/s), blue (on necks) denotes
the slower electrons ($v_F^{min}$ = 4$\times 10^7$ cm/s)
that are concentrated around the FS necks.  The free electron 
value is 1.7$\times 10^8$ cm/s.
{\it Middle panel}: Fermi surfaces with relative shift of
0.71(1,1,0) (i.e. near the point K) indicating lines of intersection.
{\it Bottom panel}: the light areas indicate the ``hot spots'' (the
intersection of the Kohn anomaly surfaces with the Fermi surface)
that are involved in strong nesting and strong coupling at
Q=0.71(1,1,0) (see
Fig. \ref{xi}).  These include the necks, and three inequivalent lines
connecting neck regions.
 }
\label{FS}
\end{figure}

The volume at 35 GPa is 51\% of that at P=0, so the conduction
electron density has doubled.  The shift in character
from $s$ to $p$ is analogous
to the $s\rightarrow d$ crossover in the heavier alkali 
metals.\cite{mcmahan}  
The occupied bandwidth 
increases by only 
14\%, much less than the free electron value 
$2^{2/3}$-1 = 59\%;
this discrepancy is accounted for by the 55\% increase in the
k=0 band mass ($m_b/m$=1.34 at P=0 to $m_b/m$=2.08 at 35 GPa).  
At P=0 in the fcc phase the FSs
are significantly nonspherical and just touch at the L points
of the BZ; necks (as in Cu), where the $p$ character is strongest,
grow with increasing pressure, and the
FS at 35 GPa is shown in Fig. \ref{FS}, colored by the
Fermi velocity.  The topology of the FS 
plays a crucial role in the superconductivity of Li, as we discuss below.

The coupling strength $\lambda$ is the
average of mode coupling constants\cite{allen}
\begin{eqnarray}
\lambda_{\vec Q\nu}&=&
  \frac{2N_{\nu}}{\omega_{\vec Q\nu}N(0)}
  \frac{1}{N}\sum_k |M^{[\nu]}_{k,k+Q}|^2
     \delta(\varepsilon_k)\delta(\varepsilon_{k+Q}),
\end{eqnarray}
with magnitude determined by the EP matrix
elements $M^{[\nu]}_{k,k+Q}$ and the nesting function $\xi(Q)$ describing the
phase space for electron-hole scattering across the
FS (E$_F$=0),
\begin{equation}
 \xi(Q)=
 \frac{1}{N} \sum_k \delta(\varepsilon_k)\delta(\varepsilon_{k+Q})
 \propto \oint\frac{d{\cal L}_k}{|\vec v_k \times
     \vec v_{k+Q}|}. 
\label{XiEqn}
\end{equation}
Here the integral is over the line of intersection of the FS and
its image displaced by $Q$, $\vec v_k \equiv \nabla_k 
\varepsilon_k$ is the
velocity, and N(0) is the FS density of states.  
Evidently $\xi(Q)$ gets large if one of
the velocities gets small, or if the two velocities become collinear.

\begin{figure}
\rotatebox{-00}{\resizebox{7cm}{8cm}{\includegraphics{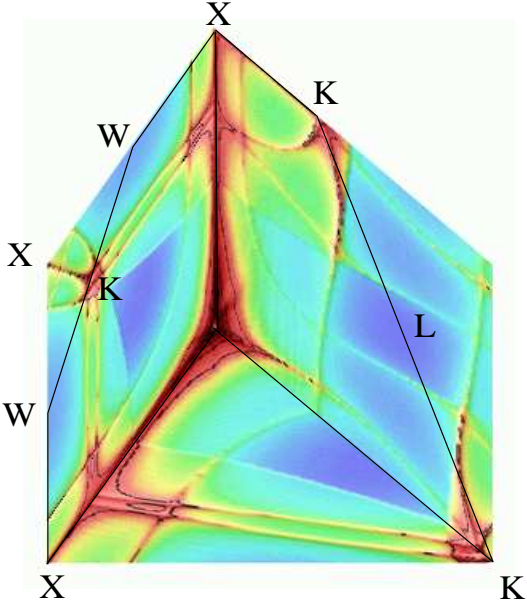}}}
\caption{(color online) Surface plots 
of the nesting function $\xi(Q)$ at 35 GPa
throughout three symmetry planes: (010) $\Gamma$-X-W-K-W-X-$\Gamma$;
(001) $\Gamma$-K-X-$\Gamma$; (110) $\Gamma$-K-L-K-X-$\Gamma$.  The
$\Gamma$ point lies in the back corner.  The dark (red) regions denote
high intensity, the light (blue) regions denote low intensity. The
maxima in these planes occur near K and along $\Gamma$-X.  To obtain the fine
structure a cubic k mesh of ($2\pi/a$)/160 was used (2$\times 10^6$
points in the BZ).
 }
\label{xi}
\end{figure}

Note that $\frac{1}{N}\sum_Q \xi(Q)$ = [N(0)]$^2$; the topology of the
FS simply determines how the fixed number of scattering processes is
distributed in Q.  For a spherical FS $\xi(Q)\propto \frac{1}{|Q|}
\theta(2k_F-Q)$; in a lattice it is simply a reciprocal lattice sum
of such functions.  This simple behavior (which would hold for bcc
Li at P=0, for example) is altered dramatically in fcc Li, as 
shown in Fig. \ref{xi} for P=35 GPa (the  
nonphysical
and meaningless $\frac{1}{|Q|}$ divergence around $\Gamma$ should
be ignored).  
There is very fine structure in $\xi(Q)$ that demands a
fine k mesh in the BZ integration,
evidence that there is 
strong focusing of
scattering processes around the K point, along the $\Gamma$-X line
peaking at $\frac{3}{4}$ $\Gamma$-X$\equiv$X$_K$, and 
also a pair of ridges (actually, cuts through surfaces) 
running in each (001) plane in K-X$_K$-K-X$_K$-K-X$_K$-K
squares.  Some additional structures 
are the simple discontinuities mentioned above, arising
from the spherical regions of the FS.  

Structure in $\xi(Q)$ arises where the integrand in Eq. \ref{XiEqn}
becomes singular, i.e. when the velocities at $k$ and $k+Q$ 
become collinear.  The FS locally is either parabolic or
hyperbolic, and the nature of the singularity is governed by the
difference surface which also is either parabolic or hyperbolic.
In the parabolic case (such as two spheres touching) $\xi(Q)$ has
a discontinuity.  In the hyperbolic case, however, $\xi(Q)$ {\it diverges}
logarithmically.  Such divergent points are not isolated, but
locally define a surface of such singularities (or discontinuities,
in the parabolic case).  The ridges and steps visible in Fig.
\ref{xi} are cuts through these singular surfaces (more details
will be published elsewhere); the intensity at K arises from transitions
from one neck to (near) another neck and is enhanced by the low neck
velocity.  Roth {\it et al.} have pointed out
related effects on the susceptibility\cite{roth} 
(which will analogously impact the real part of
the phonon self-energy), and Rice and Halperin\cite{rice} have 
discussed related processes for the tungsten FS.  In the susceptibility
(and hence in the phonon renormalization)
only FS nesting with antiparallel velocities gives rise to 
Q-dependent structure.  This explains why the ridge in $\xi(Q)$ along the
$\Gamma$-X line (due to transitions between necks and the region
between necks) does not cause much softening (see below); there will however
be large values of $\lambda_{Q\nu}$ because its structure
depends only on collinearity.

Divergences of $\xi(Q)$, which we relate
to specific regions of the FS shown in the bottom panel of Fig. \ref{FS} 
(mostly distinct from the flattened regions 
between necks discussed elsewhere\cite{rodriguez}),
specify the Q regions of greatest instability.
However, instabilities in harmonic
approximation ($\omega_{Q\nu}
\rightarrow$ 0) may not correspond to physical 
instabilities: as the frequency softens, atomic
displacements increase and the lattice can be stabilized to even
stronger coupling (higher pressure) by anharmonic interactions.
Thus, although we obtain a harmonic instability at Q$\sim$K
already at 25 GPa, it is entirely feasible that the system is
anharmonically stabilized beyond this pressure.  We infer
that indeed the regime beyond 25 GPa is an example of anharmonically
stabilized ``high T$_c$'' superconductivity.

\begin{figure}
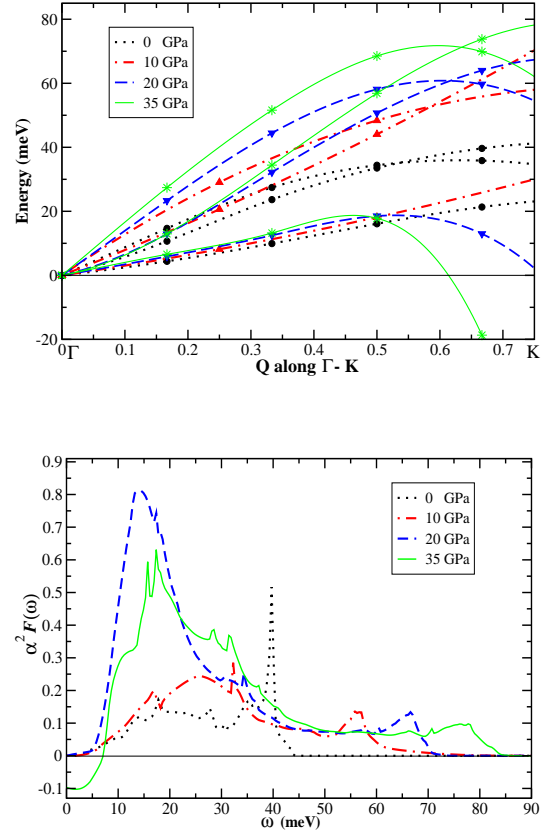

\rotatebox{-00}{\resizebox{7cm}{5cm}{\includegraphics{Fig3a.eps}}}
\vskip 10mm
\rotatebox{-00}{\resizebox{7cm}{5cm}{\includegraphics{Fig3b.eps}}}
\caption{(color online) 
{\it Top panel}: Calculated phonon spectrum (interpolated smoothly between
calculated points (solid symbols) of fcc Li along the $\Gamma$-K
direction, at the four pressures indicated.  The ${\cal T}_1$ 
(lowest) branch becomes harmonically unstable around K just above
20 GPa.   
{\it Bottom panel}: calculated spectral functions $\alpha^2 F$
for P = 0, 10, 20, and 35 GPa.  Note that, in spite of the 
(expected) increase in the maximum phonon frequency, the dominant
growth in weight occurs in the 10-20 meV region.
 }
\label{phonons}
\end{figure}

The phonon energies and
EP matrix elements have been obtained from linear
response theory as implemented in Savrasov's
full-potential linear
muffin-tin orbital code.\cite{Sav}  Phonons are calculated
at 72 inequivalent Q points (a 12$\times$12$\times$12 grid),
with a 40$\times$40$\times$40 grid
for the zone integration.
To illustrate the evolution with pressure, we use the fcc lattice
constants 8.00, 7.23, 6.80, and 6.41 bohr, corresponding approximately
to 0, 10, 20, and 35 GPa respectively (and we use these pressures
as labels).

The phonon spectrum along $\Gamma$-X behaves fairly normally.  The
longitudinal (${\cal L}$) branch at X hardens from 45 meV to 87 meV
in the 0-35 GPa
range, while the transverse (${\cal T}$) mode at X remains
at 30-35 meV.
Along $\Gamma$-L the behavior is somewhat more
interesting:
again the ${\cal L}$ branch hardens as expected, from 40 to 84 meV,
but the ${\cal T}$ branch remains
low at 15-17 meV at the
L point and acquires a noticeable dip near the midpoint at 35 GPa.
The important changes occur along
the (110) $\Gamma$-K direction as shown in Fig. \ref{phonons}:
the ${\cal L}$ and ${\cal T}_2$ branches harden
conventionally, but the $<1{\bar 1}0>$ polarized
${\cal T}_1$ branch softens dramatically
around the K point, becoming unstable around 25 GPa.  At 35 GPa
this mode is severely unstable in a substantial volume near 
the K point (not only along the $\Gamma$-K line).

We have evaluated the EP spectral function $\alpha^2 F(\omega)$
using our mesh of 72 Q points and the tetrahedron method.  Due to
the fine structure in $\xi(Q)$ and hence in $\lambda_{Q\nu}$,
numerically accurate results cannot be expected, 
but general trends should
be evident.  The resulting spectra are displayed in Fig.
\ref{phonons}(b) for each of the four pressures, showing the hardening
of the highest frequency ${\cal L}$ mode with pressure 
(43 meV $\rightarrow$ 83 meV).
The most important change is the growth in weight centered at
25 meV (10 GPa) and then decreasing to 15 meV (20 GPa) 
beyond which the instability
renders any interpretation at 35 GPa questionable.  The
growing strength is at low energy; note however that this region is
approaching the energy $\omega_{opt} = 2\pi k_B T_c 
\approx$ 10 meV which Bergmann and Rainer\cite{rainer}   
found from calculation of $\delta T_c/\delta \alpha^2 F(\omega)$
to be the optimal position to concentrate the spectral weight.
These $\alpha^2F$ spectra give the values of $\omega_{log}$,
$<\omega^2>^{1/2}$, and $\lambda$ given in Table \ref{table}.
The commonly chosen value $\mu^* = 0.13$ in the 
Allen-Dynes equation\cite{alldyn} 
(which describes the large $\lambda$ regime correctly)
gives observable values of T$_c$ = 0.4-5 in the 1-10 GPa range, but
Li is not fcc at these pressures.  The 20 K obtained for 20 GPa 
is satisfyingly close to the range of observed T$_c$, and could be
depressed to the observed value by anharmonic interactions or by
a larger value of $\mu^*$.

\begin{table}[b]
\caption{From the calculated $\alpha^2F(\omega)$ at three pressures (GPa),
the logarithmic and second moments of the frequency (K), the value of
$\lambda$, and T$_c$ (K) calculated using $\mu^*$=0.13.}
\begin{center}
\begin{tabular}{|c|c|c|c|c|}
\hline
Pressure & $\omega_{log}$ & $<\omega^2>^{1/2}$ & $\lambda$ & T$_c$  \\
\hline\hline
~0  &   209  &  277  &   0.40  &  0.4 \\
10  &   225  &  301  &   0.65  &  5~~ \\
20  &   ~81  &  176  &   3.1~  & 20~~ \\
\hline
\end{tabular}
\end{center}
\label{table}
\end{table}

We have shown how Fermi surface topology 
can concentrate scattering processes into specific surfaces in Q-space,
and even in alkali metals can lead to very strong coupling to
phonons with these momenta, and can readily drive lattice instability.
To enhance $\lambda$, it is necessary in addition that the large
regions of $\xi(Q)$ are accompanied by large EP matrix elements.
We have verified that the Q=$(\frac{2}{3},\frac{2}{3},0)
\frac{2\pi}{a}$ ${\cal T}_1$ (unstable) phonon
(near K) causes large band shifts with atomic
displacement 
($\delta\varepsilon_k/\delta u \approx$ 
5 eV/\AA) near the FS necks,
while for the stable ${\cal T}_2$ mode band shifts are
no more than 5\% of this value.  Thus the focusing of scattering processes
is indeed coupled with large, polarization-dependent matrix elements.

This focusing of EP coupling strength makes accurate 
evaluation of the total coupling strength $\lambda$ numerically
taxing.
The richness and strong $\vec Q$-dependence of the 
electron-phonon coupling that we have uncovered
may explain the overestimates of T$_c$ in the previous
work in Li, and may apply to the overestimates in boron\cite{boron}.  It is
clear however that it is EP coupling and not Coulomb 
interaction\cite{jansen} that is responsible for the impressively
high T$_c$.
Compressed Li thus has several similarities to MgB$_2$ --
very strong coupling to specific phonon modes, T$_c$ determined by
a small fraction of phonons -- but the physics is entirely
different since there are no strong covalent bonds and it is
low, not high, frequency modes that dominate the coupling.
Compressed Li is yet another system that demonstrates that our
understanding of superconductivity arising from``conventional'' 
EP coupling is far from complete, with different systems
continuing to unveil unexpectedly rich physics.

We acknowledge important communication with K. Koepernik, A. K. McMahan,
and S. Y. Savrasov.
This work was supported
by National Science Foundation grant Nos. DMR-0421810 and DMR-0312261.  
A.L. was supported
by the SEGRF program at LLNL, J.K. was supported by
DOE grant FG02-04ER46111, and H.R. was supported by DFG 
(Emmy-Noether-Program).


\begin{thebibliography}{10}

\bibitem{schilling}S. Deemyad and J. S. Schilling, Phys. Rev. Lett.
  {\bf 91}, 167001 (2003).
\bibitem{shimizu}K. Shimizu {\it et al.}, Nature {\bf 419}, 597 (2002).
\bibitem{struzhkin}V. V. Struzhkin {\it et al.}, Science {\bf 298},
  1213 (2002).
\bibitem{akimitsu} J. Nagamatsu {\it et al.} Nature {\bf 410},
  63 (2001).
\bibitem{finns}K. I. Juntunen and J. T. Tuoriniemi, Phys. Rev. Lett.
  {\bf 93}, 157201 (2004).

\bibitem{allenLi}P. B. Allen and M. L. Cohen, Phys. Rev. {\bf 187},
  525 (1969).
\bibitem{jarlborg}T. Jarlborg, Phys. Scr. {\bf 37}, 795 (1988).
\bibitem{novikov}N. E. Christensen and D. L. Novikov, Phys. Rev. Lett.
  {\bf 86}, 1861 (2001).
\bibitem{liu}A. Y. Liu {\it et al.}, Phys. Rev. B {\bf 59}, 4028 (1999).

\bibitem{neaton}J. B. Neaton and N. W. Ashcroft, Nature {\bf 400},
  141 (1999).
\bibitem{iyakutti}K. Iyakutti and C. Nirmala Louis, Phys. Rev. B
  {\bf 70}, 132504 (2004).
\bibitem{hanfland}M. Hanfland, K. Syassen, N. E. Christensen, and D.
  L. Novikov, Nature {\bf 408}, 174 (2000); N. E. Christensen and
  D. L. Novikov, Phys. Rev. Lett. {\bf 86}, 1861 (2001).

\bibitem{rodriguez}A. Rodriguez-Prieto and A. Bergara, Proc. of the
  Joint 20th AIRAPT-43rd EHPRG 2005 Conference (unpublished);
  cond-mat/0505619.

\bibitem{ucd}J. An and W. E. Pickett, Phys. Rev. Lett. {\bf 86},
  4366 (2001).
\bibitem{kortus}J. Kortus {\it et al}, Phys. Rev. Lett. {\bf 86},
  4656 (2001).
\bibitem{kong}Y. Kong {\it et al.}, Phys. Rev. B {\bf 64}, 020501 (2001).

\bibitem{allen}P. B. Allen, Phys. Rev. B {\bf 6}, 2577 (1972);
 P. B. Allen and M. L. Cohen, Phys. Rev. Lett. {\bf 29}, 1593 (1972).

\bibitem{mcmahan}A. K. McMahan, Phys. Rev. B {\bf 29}, 5982 (1984).

%

\bibitem{roth}L. M. Roth, J. J. Zieger, and T. A. Kaplan,
  Phys. Rev. {\bf 149}, 519 (1965).
\bibitem{rice}T. M. Rice and B. I. Halperin, Phys. Rev. B {\bf 1}, 
  509 (1970).
\bibitem{Sav}S. Y. Savrasov, Phys. Rev. B {\bf 54}, 16470 (1996);
 S. Y. Savrasov and D. Y. Savrasov, {\it ibid.} {\bf 54}, 16487 (1996).

\bibitem{rainer}G. Bergmann and D. Rainer, Z. Physik A {\bf 252},
  174 (1972).
\bibitem{alldyn}P. B. Allen and R. C. Dynes, Phys. Rev. B
  {\bf 12}, 905 (1975).
\bibitem{boron}S. K. Bose, T. Kato, and O. Jepsen, cond-mat/0507283.
\bibitem{jansen}L. Jansen, Physica A {\bf 332}, 249 (2004).

\end{thebibliography}
\end{document}